# Contests with a constrained choice set of effort

Doron Klunover[a], John Morgan[b]

September 2020 (final version)

**Abstract**

We consider a symmetric two-player contest, in which the choice set of effort is constrained. We apply a fundamental property of the payoff function to show that, under standard assumptions, there exists a unique Nash equilibrium in pure strategies. It is shown that all equilibria are near the unconstrained equilibrium. Perhaps surprisingly, this is not the case when players have different prize evaluations.

---

[a] Corresponding author: Department of Economics, Ariel University, 40700 Ariel, Israel. E-mail address: doronkl@ariel.ac.il.

[b] Haas School of Business and Department of Economics, University of California, Berkeley, California 94720.

## 1. Introduction

The standard assumption in contest theory that players have a convex choice set may be reasonable when units of effort are sufficiently small, but there are many important contexts in which this does not hold. For instance, it may be possible for a country to contain an overseas crisis using a small number of aircraft, but getting them there and maintaining them may require the use of an aircraft carrier. Thus, the actual choice is between a large operation and no operation at all. Likewise, there may be a large fixed cost in supplementing air power with "boots on the ground". More generally, the choice set may be further constrained by rules, laws and regulations, technological constraints, political constraints and the like. Our model extends the contest literature, which usually assumes away problems of "granularity" of effort, or alternatively, imposes a particular constraint, such as entry costs fees (Moldovanu and Sela, 2001), caps on expenditure (Che and Gale, 1998), budget constraints (Kvasov, 2007), a discrete strategy space (Baye et al.,1994), etc.[1] In particular, we consider a contest between two identical risk-neutral players and a contest success function (CSF) with a concave impact function, in which the choice set of effort can be (almost) any subset of the real line.

We apply a fundamental property of payoff functions in contests in order to show that a Nash equilibrium in pure strategies exists and, except in a special case, is unique. Specifically, equilibrium effort is one of the two feasible elements that are closest to the unconstrained equilibrium. We note that in some circumstances small changes in the strategy space produce large changes in effort.

[1] For a review of the literature, see Konrad (2009), Corchón and Serena (2018), Fu and Wu (2019).

Although we focus on equilibrium in pure strategies and do not fully characterize the set of equilibria in mixed strategies, we show that such equilibria, if they exist, are also limited to the same two effort set and therefore are also near the unconstrained equilibrium.

We show that the intuitive results above are not obtained when players have different prize evaluations. For this case, we present a counterexample, in which the choice set is first fixed such that the equilibrium effort is not close to the unconstrained equilibrium, and then fixed such that a Nash equilibrium in pure strategies does not exist.

The rest of the paper is structured as follows. In section 2, we describe the model. Sections 3 characterizes equilibria in pure strategies and discusses the case of mixed strategies. An example in which players have different prize evaluations is presented in section 4. Section 5 concludes.

## 2. The model

Two risk-neutral players compete for a prize with a common value $v$. Each player $i \in \{1,2\}$ chooses an irreversible non-negative effort $e_i$ from the choice set $S \subseteq [0,\infty)$.

The probability of player $i \in \{1,2\}$ winning the prize is determined by the following CSF:

$$(1)\ p(e_i, e_j) = \begin{Bmatrix} \frac{f(e_i)}{f(e_i)+f(e_j)}, & \text{if } (e_i, e_j) \neq (0,0) \\ \frac{1}{2}, & \text{if } (e_i, e_j) = (0,0) \end{Bmatrix},$$

where $f'(\cdot) > 0, f''(\cdot) \leq 0$ and $f(0) = 0$.

We therefore assume that $f(e)$ (usually referred to as the player's impact function) is twice differentiable and concave. The CSF in (1) has been axiomatized and is commonly used in the literature (see Skaperdas, 1996). Furthermore, Baye and Hoppe (2003) show that a contest with a CSF which is a special case of (1) and a discrete strategy space is strategically equivalent to innovation tournaments and patent-race games.[2]

Player $i's$ problem is therefore:

$$(2)\ \max_{e_i} E\pi(e_i; e_j) \text{ such that } e_i \in S \text{, where } E\pi(e_i, e_j) = \frac{f(e_i)}{f(e_i) + f(e_j)} v - e_i \quad \forall i \in \{1,2\}.$$

For the case in which $S = [0,\infty)$, let $Ri(e_j)$ be player $i$'s best response function.

Yildirim (2005) (who builds on Dixit, 1987) then notes the following important facts.

**Fact 1** Given $e_j$, $E\pi(e_i; e_j)$ is strictly concave in $e_i$. (Yildirim 2005, Fact 1)

**Fact 2** $Ri(e_j)$ is strictly increasing if the pair $(e_i,e_j)$ is such that $e_i>e_j$; it reaches its maximum if $e_i=e_j$; and it is strictly decreasing if $e_i<e_j$. (Yildirim 2005, Fact 2)

[2] Note that Baye and Hoppe (2003) focus on showing strategic equivalence rather than analyzing equilibrium behavior; therefore, for the case of non-increasing return on effort they refer the reader to the well-known unconstrained equilibrium.

As noted by Yildirim (2005), Fact 2 implies that if $S = [0,\infty)$, then there is a unique symmetric interior solution which we denote by $e^*$.[3]

Let $\underline{e}$ and $\bar{e}$ be the elements in $S$ that are the closest (from below and from above) to $e^*$. Formally, if $e^*$ is not in $S$, then $\nexists e \in S | e \in (\underline{e}, \bar{e})$, where $0 \leq \underline{e} < e^* < \bar{e}$ and $\underline{e}$ and $\bar{e}$ are in $S$. If $e^*$ is in $S$, then $\underline{e} \equiv e^* \equiv \bar{e}$.

In the remainder of the paper, we assume that the elements $\underline{e}$ and $\bar{e}$ exist. Otherwise, equilibrium in pure strategies may not exist.

## 3. Nash Equilibria

In the following, we focus on pure strategies and later in this section we briefly discuss the case in which mixed strategies are allowed.

First, pursuant to (2), we present the following useful and perhaps important property of the player's payoff function:

$$(3)\ E\pi(x,y) - E\pi(y,y) = p(x,y)v + y - \left(\frac{v}{2} + x\right)$$

$$= \frac{v}{2} - x - (v(1 - p(x,y)) - y) = E\pi(x,x) - E\pi(y,x).$$

As far as we know, this result is presented here for the first time and may have wider applications beyond the scope of this paper. For instance, it implies that in a contest where choice is limited to a set of two effort levels, there always exists a dominant strategy.

[3] Notice that the facts in Yildirim (2005) are more general since they refer also to the case in which $f_i(e) \neq f_j(e)$.

Building on (3) and Facts 1 and 2, we introduce Lemmas 1 and 2 which present important equilibrium features and some additional technical results needed for the rest of the analysis. Following that, Proposition 1 characterizes the complete equilibria in pure strategies.

***Lemma 1*** *Each player's equilibrium effort is an element from the set* $\{\underline{e}, \overline{e}\}$.

All proofs appear in the Appendix.[4] Note that Lemma 1 implies that if $e^*$ is in $S$, then it is the unique equilibrium. More generally, while Lemma 1 significantly reduces the candidates for equilibrium strategies, it remains to characterize the resulting set of equilibria. It proves useful to define a *threshold* effort level $\hat{e}$ with the following properties: (1) When her rival chooses $\overline{e}$, a player is indifferent between choosing the threshold and $\overline{e}$. (2) Likewise, when her rival chooses $\hat{e}$, a player is again indifferent between choosing $\hat{e}$ and $\overline{e}$. Formally, $\hat{e}$ solves $E\pi(\hat{e}, e') = E\pi(\overline{e}, e')$ for $e' \in \{\hat{e}, \overline{e}\}$. The expression $\hat{e}$ is well-defined as long as $\overline{e}$ is not too large. Otherwise, no interior solution exists for solving the indifference condition. Lemma 2 formally establishes the properties of the threshold.

***Lemma 2*** *If* $\overline{e} \leq v/2$, *then there exists a unique* $\hat{e} \in [0, e^*]$ *which is monotonically decreasing in* $\overline{e}$*; otherwise* $\nexists \hat{e}$.

Note that Lemma 2 implies that the upper bound of $\overline{e}$ for which there exists $\hat{e}$, is independent of the particular structure of the impact function $f(e)$. In Proposition 1, we characterize the equilibria structure.

---

[4] The proof for Lemma 1 considers only pure strategies. However, in a separate proof in the end of the Appendix, we show that Lemma 1 holds also when mixed strategies are allowed.

***Proposition 1*** *The Nash equilibria in pure strategies are as follows:*

a. *Effort is* $\underline{e}$ *when* $\bar{e}>v/2$ *or* $\hat{e}<\underline{e}$.
b. *Effort is* $\bar{e}$ *when* $\hat{e}>\underline{e}$.
c. *Any pair of efforts* $(e_1,e_2)$ *such that* $e_i \in \{\underline{e},\bar{e}\}$ *for all* $i\in\{1,2\}$ *is an equilibrium when* $\hat{e}=\underline{e}$.

An interesting property of the equilibria presented in Proposition 1 is that small changes in the strategy space can produce large changes in effort when $\underline{e}$ is close to $\hat{e}$. This is illustrated in the following simple example:

**Example 1** Let $S=\{0\}\cup[v/2+\varepsilon,v]$, where $\varepsilon>0$. Then, there exists a unique effortless equilibrium. However, if we let $\varepsilon<0$ (such that $v/2+\varepsilon\geq e^*$), then the player exerts $v/2+\varepsilon$ and there is almost full rent dissipation.

Note that Lemma 1 implies that any equilibrium in mixed strategies, if it exists, must also be close to $e^*$. Specifically, such an equilibrium must be a convex combination of $\underline{e}$ and $\bar{e}$ for which the player's best responses are $\underline{e}$ and $\bar{e}$. In this note, we focus on equilibria in pure strategies and therefore leave the analysis of mixed strategies equilibria for future research.

## 4. Players with different prize evaluations

In this section, we present an example in which the CSF is a special case of (1) and the choice set is constrained, but players have different prize evaluations. First, we fix $S$ such that the equilibrium effort of one of the players is *not* close to her unconstrained equilibrium effort, and then we consider a different $S$ for which an equilibrium in pure strategies does not exist.

**Example 2** Assume that $f(e)=e$ and $S = [0,\infty)$, and let the valuations of the prize by players 1 and 2 be 1 and 2, respectively. The equilibrium effort is then (2/9, 4/9).

Now, assume instead that $S=\{9/50, 1/5, 5/9\}$. Then the matrix of the players' payoffs is:

| player 2/ player 1 | 9/50 | 1/5 | 5/9 |
|---|---|---|---|
| 9/50 | 0.32, 0.82 | 0.293, 0.852 | 0.06471, 0.955 |
| 1/5 | 0.326, 0.767 | 0.3, 0.8 | 0.0647, 0.915 |
| 5/9 | 0.199, 0.309 | 0.179, 0.329 | -0.055, 0.444 |

Equilibrium is then (9/50, 5/9) and therefore player 1's equilibrium effort violates Lemma 1.

Now assume instead that $S=\{1/9, 1/5, 2/3\}$. Then the matrix of the players' payoffs becomes:

| player 2/ player 1 | 1/9 | 1/5 | 2/3 |
|---|---|---|---|
| 1/9 | 0.388, 0.888 | 0.246, 1.085 | 0.031, 1.047 |
| 1/5 | 0.442, 0.603 | 0.3, 0.8 | 0.03, 0.871 |
| 2/3 | 0.19, 0.174 | 0.102, 0.261 | -0.16, 0.33 |

A Nash equilibrium in pure strategies therefore does not exist.

## 5. Conclusion

We characterized Nash equilibria in pure strategies in a class of symmetric two-player contests in which the assumption of a convex strategy space has been relaxed.

**Acknowledgements**

We thank an anonymous referee for valuable comments. In addition, this work has benefitted from useful discussions with Nirit Agay, Christian Ewerhart, Arthur Fishman, Aart Gerritsen, Kai Konrad, Dan Kovenock, Ella Segev, Marco Serena, and Huseyin Yildirim. We also thank the audiences at the Public Choice Society meeting, the Contests: Theory and Evidence Conference, the AEA annual meeting, and a seminar held at the Max Planck Institute of Tax Law and Public Finance. All remaining errors are those of the authors. Doron Klunover thanks the Max Planck Institute of Tax Law and Public Finance for its support and hospitality (part of this paper was written during his visit there).

**Appendix:**

Proof of Lemma 1 when mixed strategies are not allowed: Fact 2 implies that:

$(A.1)\ e^* = maxR_i(e_j) = R_i(e^*)$.

By Fact 1 and (A.1), in any equilibrium, effort $e$ satisfies:

$(A.2)\ e \leq \bar{e}$.

In what follows, we show that equilibrium effort cannot be smaller than $\underline{e}$.

Fact 1 and Fact 2 imply that:

$$(A.3)\ E\pi(x,x) > E\pi(y,x)\ \forall\ e^* \geq x > y \geq 0.$$

Applying (3) to (A.3) results in:

$$(A.4)\ E\pi(x,y) > E\pi(y,y)\ \forall\ e^* \geq x > y \geq 0.$$

Let $\underline{e} \geq x \geq y \geq 0$. By (A.3) and (A.4), $(x,y)$ can be an equilibrium only when $y$=$x$=$\underline{e}$. Moreover, let $\bar{e} > e^*$. If $y$<$\underline{e}$, then $(\bar{e},y)$ cannot be an equilibrium either. This can be shown by contradiction: Assume for now that $(\bar{e},y)$ is an equilibrium. Then, by (3), $E\pi(y,y) = E\pi(\bar{e},y)$, but by (A.4), $E\pi(y,y) < E\pi(\underline{e},y)$ and therefore $E\pi(\underline{e},y) > E\pi(\bar{e},y)$, which contradicts the initial assumption that $(\bar{e},y)$ is an equilibrium. Thus, any $e$ in $S$ such that $e$<$\underline{e}$ cannot be an equilibrium effort. QED

Proof of Lemma 2: By (1), (2) and (3),

$$(A.5)\ E\pi(e,\bar{e}) = E\pi(\bar{e},\bar{e})$$

$$\leftrightarrow$$

$$E\pi(e,e) = E\pi(\bar{e},e)$$

$$\leftrightarrow$$

$$\frac{v}{2} - e = p(\bar{e},e)v - \bar{e}.$$

Let $e \leq e^*$. By definition, (A.5) is satisfied at $\bar{e} = e^*$ iff $e = e^*$, while by (1), (A.5) is satisfied at $e$=0 iff $\bar{e} = v/2$. Therefore, since Fact 1 and Fact 2 imply that the RHS of the third equality in (A.5) is decreasing in $\bar{e}(\geq e^*)$, while the LHS is independent of $\bar{e}$, for any given $\bar{e} \in (e^*, v/2)$, at $e$=0 ($e = e^*$) the LHS of the third expression of (A.5) is smaller (greater) than the RHS. Therefore, given that both sides of the third equality of (A.5) are decreasing in $e$ but the LHS is linear and by (1) the RHS is convex, by the Intermedium Value-theorem, for any given $\bar{e} \in (e^{*}, v/2)$, there exists a unique $\hat{e}$ over the interval $(0, e^{*})$ such that $e \gtreqless \hat{e} \leftrightarrow E\pi(e,e) \gtreqless E\pi(\bar{e}, e)$, where $\hat{e}$ is monotonically decreasing in $\bar{e}$ .[5]     QED

Proof of Proposition 1: By (3), the player's equilibrium effort in the symmetric 2x2 contest where choice is limited to $\{\underline{e}, \bar{e}\}$ is determined by the following inequality:

$$\text{(A.6)}\ E\pi(\underline{e}, \bar{e}) \gtreqless E\pi(\bar{e}, \bar{e})$$

$$\leftrightarrow$$

$$E\pi\big(\underline{e}, \underline{e}\big) \gtreqless E\pi\big(\bar{e}, \underline{e}\big).$$

According to the proof of Lemma 2, if $\bar{e} > v/2$, then $E\pi\big(\underline{e}, \underline{e}\big) > E\pi\big(\bar{e}, \underline{e}\big)$ for all $\underline{e}$; otherwise $\underline{e} \gtreqless \hat{e} \leftrightarrow E\pi\big(\underline{e}, \underline{e}\big) \gtreqless E\pi\big(\bar{e}, \underline{e}\big)$. Lemma 1 implies that the set of equilibria in the original contest, in which the choice set is $S$, is a subset of the set of equilibria in the 2x2 contest. In what follows, it is shown that these two sets coincide.

If $(\underline{e}, \underline{e})$ is an equilibrium in the 2x2 contest, then $E\pi\big(\underline{e}, \bar{e}\big) \geq E\pi(\bar{e}, \bar{e})$ which by Fact 1 and (A.1) implies that $E\pi\big(\underline{e}, \bar{e}\big) \geq E\pi(\bar{e}, \bar{e}) > E\pi(y, \bar{e})$ for all $y > \bar{e}$. Therefore, by (A.3), if $(\underline{e}, \underline{e})$ is an equilibrium in the 2x2 contest, then it is an equilibrium in the original contest. In addition, if $(\bar{e}, \bar{e})$ is an equilibrium in the 2x2 contest, then $E\pi(\bar{e}, \bar{e}) \geq E\pi\big(\underline{e}, \bar{e}\big)$ which by Fact 1 implies that $E\pi(\bar{e}, \bar{e}) \geq E\pi\big(\underline{e}, \bar{e}\big) > E\pi(y, \bar{e})$ for all $y < \underline{e}$. Therefore, by (A.1), if $(\bar{e}, \bar{e})$ is an equilibrium in the 2x2 contest, then it is in the original contest as well.     QED

[5] Note that it also implies that $v/2 - e > p(\bar{e}, e)v - \bar{e}$ for all $e \leq e^{*}$ when $\bar{e} > v/2$.

Proof for Lemma 1 when mixed strategies are allowed: Assume that, $(qe_1+(1-q)e_2,q'e_3+(1-q')e_4)$ where $0<q,q'<1$, is an equilibrium.[6] By (A.1) and Fact 1, for all $k=1,2,3,4$, either $e_k \in \{\underline{e}, \overline{e}\}$ or $e_k<\underline{e}$.

Without loss of generality assume that: $e_1<e_2$ and $e_1<\underline{e}$. Then, by (A.3) and Fact 1, $q'e_3+(1-q')e_4>e^*$, which (without loss of generality) implies that $e_4=\overline{e}$. It follows that, $e_2\leq\underline{e}$ and $e_3=\underline{e}$; Note that $\overline{e}$ and $e(<\underline{e})$ cannot be best response to the same effort, since by Fact 1, $\underline{e}$ is always better than one of them.

Given that, $qe_1+(1-q)e_2<e_2,\underline{e}<q'\underline{e}+(1-q')\overline{e}$, $(qe_1+(1-q)e_2)\in(e_1,e_2)$ and $(q'\underline{e}+(1-q')\overline{e})\in(\underline{e},\overline{e})$, by fact 1, $E\pi(q'\underline{e}+(1-q')\overline{e},qe_1+(1-q)e_2)>E\pi(qe_1+(1-q)e_2,qe_1+(1-q)e_2)$ and $E\pi(q'\underline{e}+(1-q')\overline{e},q'\underline{e}+(1-q')\overline{e})<E\pi(qe_1+(1-q)e_2,q'\underline{e}+(1-q')\overline{e})$, which by (3) is a contradiction. QED

[6] Note that Fact 1 implies that, in equilibrium in mixed strategies, a player cannot mix over more than two (pure) strategies.